# BLOCKING TEMPERATURE OF A SYSTEM OF CORE/SHELL NANOPARTICLES

Afremov L.L.[1], Anisimov S.V.[2], Iliushin I.G.[3]

Affiliation: Far Eastern Federal University, Russia, 690091, Vladivostok, Sukhanova st. 8;

e-mail: [1] afremov.ll@dvfu.ru
[2] anisimovsv25@gmail.com
[3] iliushin.ig@dvfu.ru

**Abstract**

A theoretical study was conducted of the size dependence of the blocking temperature $T_b$ of a system of interacting core/shell nanoparticles. A method for estimating the blocking temperature of interacting core/shell nanoparticles is presented, which allows $T_b$ to be calculated more correctly than using the "Neel relation". It was shown that together with an increase in the intensity of the magnetostatic interaction (concentration of nanoparticles) the blocking temperature increases, while the growth of the external magnetic field leads to the opposite effect. Moreover, the $T_b$ of large nanoparticles changes more significantly. A comparison of different methods for determining the blocking temperature from the ZFC and FC curves showed that the method for determining $T_b$ using the temperature derivative of the difference between ZFC and FC is more preferable.

**Introduction**

The use of magnetic nanoparticles in various engineering, chemical, medical, and other technologies [for example, see ref. 1] initiated a study of the influence of the geometric, structural, and morphological properties of nanoparticles on their hysteretic characteristics and blocking temperature $T_b$. In contrast to the coercive force $H_c$, saturation magnetization $M_s$, remanent saturation magnetization $M_{rs}$, and the bias field $H_e$ characterizing the magnetization curve, the blocking temperature is a characteristic of the nonequilibrium (relaxation) behavior of a system of magnetic nanoparticles. It can be defined as the temperature above which the system of magnetic nanoparticles behaves like a paramagnetic gas, a superparamagnetic state passes. Moreover, at temperatures above blocking ($T \geq T_b$), up to the critical temperature of the magnetic phase transition $T_c$, the magnetization curve becomes reversible: $H_c = 0$ and $M_{rs} = 0$. The transition to the superparamagnetic state is associated with the superiority of the energy of thermal fluctuations $k_B T$ over the energy of potential barriers $\Delta E$, which separate the equilibrium states of the magnetic moments of nanoparticles ($k_B T > \Delta E$).

The temperature of the blocking temperature $T_b$ of the nanoparticle system is usually estimated using the relation:

$$T_b = \frac{K_{eff}(T_b)V}{k_B \ln(\tau f_0)}, \qquad (1)$$

which follows from the expression for the relaxation time $\tau$ of a system of single-phase (having a single magnetic phase) uniaxial single-domain particles with a volume V and an effective anisotropy constant $K_{eff}$ [2, 3]. In (1), it is assumed that the frequency factor $f_0 \approx 10^{10}$ s$^{-1}$ and the relaxation time is equal to the measurement time $t_{exp}$. Some authors (see, for example, [4–9]) use relation (1) to calculate $T_b$ or effective anisotropy constant and in the case of two-phase (core/shell) nanoparticles, which, as will be shown below, is completely incorrect. Note that incorrect application of (1) is associated with a multitude of relaxation times due to the spectrum of magnetic states of the magnetic moment and potential barriers separating them in a heterophase particle.

Experimentally, the average blocking temperature $\langle T_b \rangle$ is determined in various ways, by:

- the point of separation of the ZFC curve from FC [10–12],
- the maximum of the ZFC curve [5, 10, 13–22],
- the maximum derivative of the difference between the curves FC (T) - ZFC (T) [9, 10, 23–26],
- the maximum derivative of the thermo-remanent magnetization TRM (T) [9],
- the maximum derivative of the remanent magnetization $M_R$ (T) [27]
- the maximum temperature dependence of the real part of the magnetic susceptibility [3, 28].
- Mössbauer spectra comparison [4, 5, 29–31]

In this paper, in the framework of the theory of interacting core/shell nanoparticles [32], a simulation of the dependence of the blocking temperature on the sizes of nanoparticles, their magnetostatic interaction (concentration), and external magnetic field, is presented. A theoretical comparison of the main methods for determining the blocking temperature from the FC and ZFC curves is made, and the correct method for calculating the $T_b$ of the core/shell nanoparticle system is proposed.

## 1. Magnetization of a system of core/shell nanoparticles

Let us consider a system of N interacting core/shell nanoparticles uniformly distributed over the volume $V_0$. We assume that nanoparticles of volume V having the shape of elongated ellipsoids are distributed over sizes $a$ with probability $f(a)da$. In the approximation of the model described in detail in [32], the magnetization of the system of nanoparticles is determined by the following relation:

$$M(t) = \int c \left[ \left((1-\varepsilon)\mathcal{M}_s^{(1)} + \varepsilon\mathcal{M}_s^{(2)}\right) \left(n_1(t,a,h) - n_3(t,a,h)\right) + \right.$$
$$\left. + \left((1-\varepsilon)\mathcal{M}_s^{(1)} - \varepsilon\mathcal{M}_s^{(2)}\right) \left(n_2(t,a,h) - n_4(t,a,h)\right) \right] f(a)da\, W(h)dh. \quad (2)$$

Here $c = NV/V_0$ is the volume concentration of core / shell nanoparticles, $\mathcal{M}_s^{(1)}$ and $\mathcal{M}_s^{(2)}$ are the spontaneous magnetizations of the shell and core of the nanoparticle, respectively, $v$ and $\varepsilon = v/V$ are the volume and relative volume nuclei, respectively, $W(h)$ is distribution density over random interaction fields $h$, which is described in [32, 33], $n_i(t,a,h)$ are the populations of four magnetic states of nanoparticles, and in the first state the magnetic moments of the core and shell are oriented in parallel (↑↑), in the second - antiparallel (↑↓), and in the third and fourth orientation of magnetic moments is inverse to the first two: (↓↓) and (↓↑) respectively. According to [32], the populations are determined using matrix exponent

$$\boldsymbol{N}(t) = \exp\left(\widetilde{W}t\right) \cdot \boldsymbol{N_0} + \int_0^t exp\left(\widetilde{W}(t-\tau)\right)d\tau \cdot \boldsymbol{V}, \quad (3)$$

$$\boldsymbol{N}(t) = \begin{pmatrix} n_1(t) \\ n_2(t) \\ n_3(t) \end{pmatrix}, \widetilde{W}_{ik} = \begin{cases} -\sum_{j \neq i}^4 W_{ij} - W_{4i}, & i = k, \\ W_{ki} - W_{4i}, & i \neq k, \end{cases}, \boldsymbol{V} = \begin{pmatrix} W_{41} \\ W_{42} \\ W_{43} \end{pmatrix}, \quad (4)$$

$n_4(t)$ is expressed from the normalization condition: $n_1(t) + n_2(t) + n_3(t) + n_4(t) = 1$, $W_{ik} = f_0 exp(-E_{ik}/k_B T)$ are matrix elements matrices of probabilities of transition from the ith equilibrium state to the k-th, $f_0$ is frequency factor [26], $E_{ik} = E_{ik}^{(max)} - E_i^{(min)}$ is the height of the potential barrier, and $E_{ik}^{(max)}$ is the smallest of the maximum energies that correspond to the transition of the magnetic moment from the *i*-th equilibrium state with energy $E_i^{(min)}$ to k-th state. The expressions for potential barriers $E_{ik}$ are presented in [32, Appendix 2].

### 1.1. Blocking temperature of core/shell nanoparticles

According to [2, 3] when estimating the blocking temperature $T_b$ of single-phase (having one magnetic phase) single-domain particles with volume V, we use the expression for the relaxation time τ, which is expressed in terms of the frequencies of transitions $W_{ik}$ from one equilibrium state of the particle to another: $1/\tau(T_b) = W_{12}(T_b) + W_{21}(T_b)$, where $W_{ik}(T_b) = f_0 exp(-E_{ik}(T_b)/k_B T_b)$, $E_{12}$ ($E_{21}$) is a potential barrier to the transition from 1-st (2-nd) to 2-nd (1-st) state.

Two-phase (core-shell) nanoparticles, unlike single-phase ones, can be in one of four (six or eight) magnetic states [32]. Therefore, the relaxation times of nanoparticles are determined by the matrix of frequencies of transitions from the *i*-th state to the *k*-th state: $W_{ik} = f_0 exp(-E_{ik}/k_B T)$.

The spectrum of relaxation times $\tau_k$ core-shell nanoparticles is expressed in terms of the eigenvalues $w_k$ of the transition matrix $W_{ik}$

$$Det\,|W_{ik} - w_k \delta_{ik}| = 0, \qquad (5)$$

which are the inverse times of the transition from one state to another $|w_k| = 1/\tau_k$.

To estimate the time of transition to the equilibrium state, we will use the maximum of them, $\tau = \tau_{k\,max}$, realizing that transitions with shorter relaxation times are already completed. Then, to estimate the blocking temperature $T_b$ of interacting nanoparticles with size a, we can use the relation

$$\tau(T_b) = \tau_{k\,max}(T_b, a, h) = t_{exp}. \qquad (6)$$

That is, we will relate the relaxation time to "blocked" nanoparticles, which is equal to (or more) the measurement time $t_{exp}$.

In real systems, nanoparticles are distributed over sizes *a* and fields of the magnetostatic interaction *h*; therefore, in what follows, to calculate the blocking temperature, we will use relation (6) averaged over *a* and *h*:

$$\int \tau_{k\,max}(T_b, a, h)\,f(a)da\,W(h)dh = t_{exp}. \qquad (7)$$

### 2. Selection of the modeling parameters

For comparison with experimental data [8], the calculations used the geometric characteristics of core/shell $Zn_{0.5}Mn_{0.5}Fe_2O_4/Fe_3O_4$ nanoparticles studied in detail in [8]. When integrating expression (2), the law of the lognormal distribution of nanoparticle sizes *a* was used:

$$f(a) = \frac{1}{a\sqrt{2\pi\sigma^2}} exp\left(-\frac{(log[a/\langle a \rangle])^2}{2\sigma^2}\right), \qquad (8)$$

with the mean size $\langle a \rangle$ and dispersion $\sigma$ given in [8] (see Table 1).

**Table 1.** Mean size $\langle a \rangle$ and dispersion $\sigma$ of $Zn_{0.5}Mn_{0.5}Fe_2O_4/Fe_3O_4$ nanoparticles [8].

| Sample number | 1 | 2 | 3 | 4 |
|---|---|---|---|---|
| $\langle a \rangle$, nm | 10 | 11.5 | 12.2 | 13.0 |
| $\sigma$, nm | 0.6 | 0.6 | 0.6 | 0.6 |

In addition, we took into account the dependence of spontaneous magnetization and crystallographic anisotropy of $Zn_{0.5}Mn_{0.5}Fe_2O_4$ ferrite and magnetite on their size and temperature. So the spontaneous magnetizations of magnetite and ZnMn ferrite were estimated using the relations:

$$\mathcal{M}_s^{(Fe_3O_4)}(a,T) = \mathcal{M}_{s\ bulk}^{(Fe_3O_4)}(T_0)\left(1 - \frac{6t}{a}\right)\left(\frac{T_c^{(Fe_3O_4)}(a) - T}{T_c^{(Fe_3O_4)}(a) - T_0}\right)^{1/2},$$

$$\mathcal{M}_s^{(ZnMn)}(a,T) = \mathcal{M}_{s\ bulk}^{(ZnMn)}(T_0)\left(\frac{T_c^{(ZnMn)}(a) - T}{T_c^{(ZnMn)}(a) - T_0}\right)^{\frac{1}{2}},$$

where $\mathcal{M}_{s\ bulk}^{(Fe_3O_4)}(T_0) = 88{,}65$ emu/g, $t = 2{,}26$ nm [34, 35], $\mathcal{M}_{s\ bulk}^{(ZnMn)}(T_0) = 42$ emu/g, [8], $T_0$ is the room temperature.

The dependence of the Curie temperature on the size of magnetite nanoparticles was determined as follows:

$$T_c^{(Fe_3O_4)}(a) = 856\left(1 - \frac{a_1}{a}\right)^{\frac{1}{\nu}}$$

Where, according to [36] $a_1 = 0{,}51 \cdot 10^{-7}$ cm and $\nu = 0{,}82$.

The Curie temperature of the ZnMn ferrite nanoparticles was taken as $T_c^{(ZnMn)}(a) = 380$ K [37].

The constants of "bulk" K_A and surface anisotropy K_S are equal: for magnetite $K_A = -1{,}1 \cdot 10^5$ erg/cm$^3$ [31], $K_S = 2.03*10^{-2}$ erg/cm$^2$ at 4K [38], and accordingly, for ZnMn ferrite, $K_A = 3.8 \cdot 10^3$ erg/cm$^3$ [39], $K_S$ was taken as a model parameter and was chosen so as to bring the calculated value of the blocking temperature closer to one of the experimental ones. It turned out that $K_S = 0.15$ erg/cm$^2$.

### 3. Results
#### 3.1. Calculation of blocking temperature using the relaxation time

In this section, we study the dependence of the blocking temperature $T_b$ on the sizes of core/shell nanoparticles a, the intensity of their dipole–dipole interaction (concentration of nanoparticles $c$), and the external magnetic field $H$. The blocking temperature of core/shell nanoparticles $Zn_{0.5}Mn_{0.5}Fe_2O_4/Fe_3O_4$ was calculated using the expression (7).

#### 3.2. Modeling the size dependence of the blocking temperature

The dependence of the blocking temperature $T_b$ of a system of noninteracting nanoparticles on their sizes a is shown in Fig. 1. For comparison, a similar dependence of the blocking temperature of magnetite nanoparticles is given. As one would expect, in the region of large sizes of $T_b$ core-shell nanoparticles and magnetite particles coincide. In the size range from 8 nm to 13 nm, the blocking temperature varies nonmonotonously (see the inset in Fig. 1).

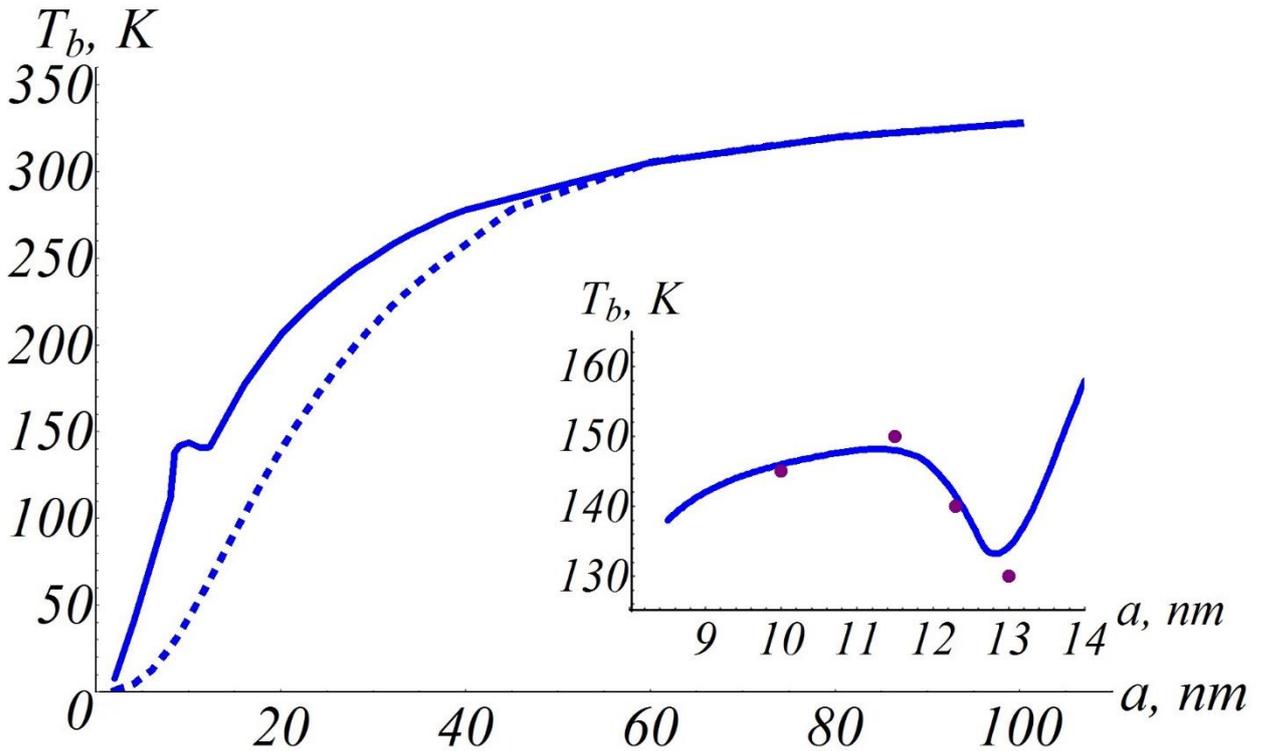

**Fig. 1.** Dependence of the blocking temperature $T_b$ on the sizes a of nanoparticles $Zn_{0.5}Mn_{0.5}Fe_2O_4/Fe_3O_4$ (solid line) and magnetite (dashed line). The inset shows the nonmonotonic size dependence of $T_b$. Dots mark the results of the experiment [8].

This is due to the peculiarities of the dependence of the potential barriers of nanoparticles $Zn_{0.5}Mn_{0.5}Fe_2O_4/Fe_3O_4$ on their sizes (see Fig. 2).

Magnetization of nanoparticles (transition to equilibrium states) is mainly determined by potential barriers $E_{ik}$ of the smallest height. The results of calculations of $E_{ik}$ carried out using the relations described in Appendix II of [32] are presented in Fig. 3. It can be seen from the figures that in the range of sizes (10 - 11.5) nm, the magnetization of the system of nanoparticles (transition to the first state in which the magnetic moments of the core and shell are parallel) is due to transitions from the fourth state to the first, since $E_{41} \leq E_{21}, E_{31}$.

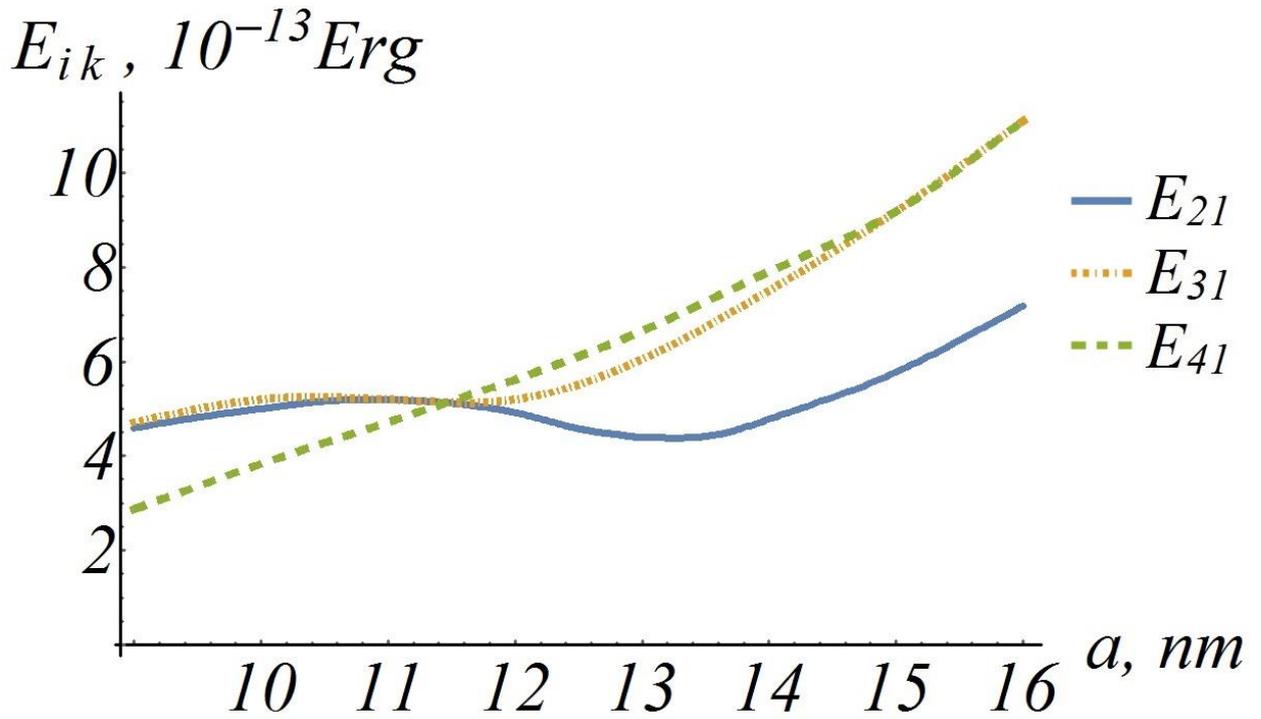

**Fig. 2.** Dependence of potential $E_{ik}$ barriers on the sizes a of $Zn_{0.5}Mn_{0.5}Fe_2O_4/Fe_3O_4$ nanoparticles at a temperature T = 145 K.

At $a > 11,5$ nm, the magnetization is determined by $E_{21}$ barriers, which change nonmonotonically with increasing nanoparticle size, reaching a minimum at $a \approx 13$ nm, which leads to a nonmonotonic change in the blocking temperature.

Note that in [32], the dependence of the blocking temperature $T_b$ on the size of nanoparticles $a$ varies according to the law $T_b(a) \sim a^3$. Such a dependence $T_b(a)$ was obtained using relation (1) under the assumption that the effective anisotropy constant does not change with increasing temperature. This approximation is not true. The temperature dependence of the anisotropy constant also leads to a deviation from the law $T_b(a) \sim a^3$ (see Fig. 1).

### 3.3. The dependence of the blocking temperature on the external field

Measurement of the blocking temperature $T_b$ from the characteristic temperature dependence of a particular magnetization involves specifying a certain value of the external magnetic field H. This field, changing potential barriers separating the equilibrium states of nanoparticles, can also change $T_b$. In fig. Figure 3 shows the dependence of the blocking temperature on the H of a system of noninteracting nanoparticles $Zn_{0.5}Mn_{0.5}Fe_2O_4/Fe_3O_4$.

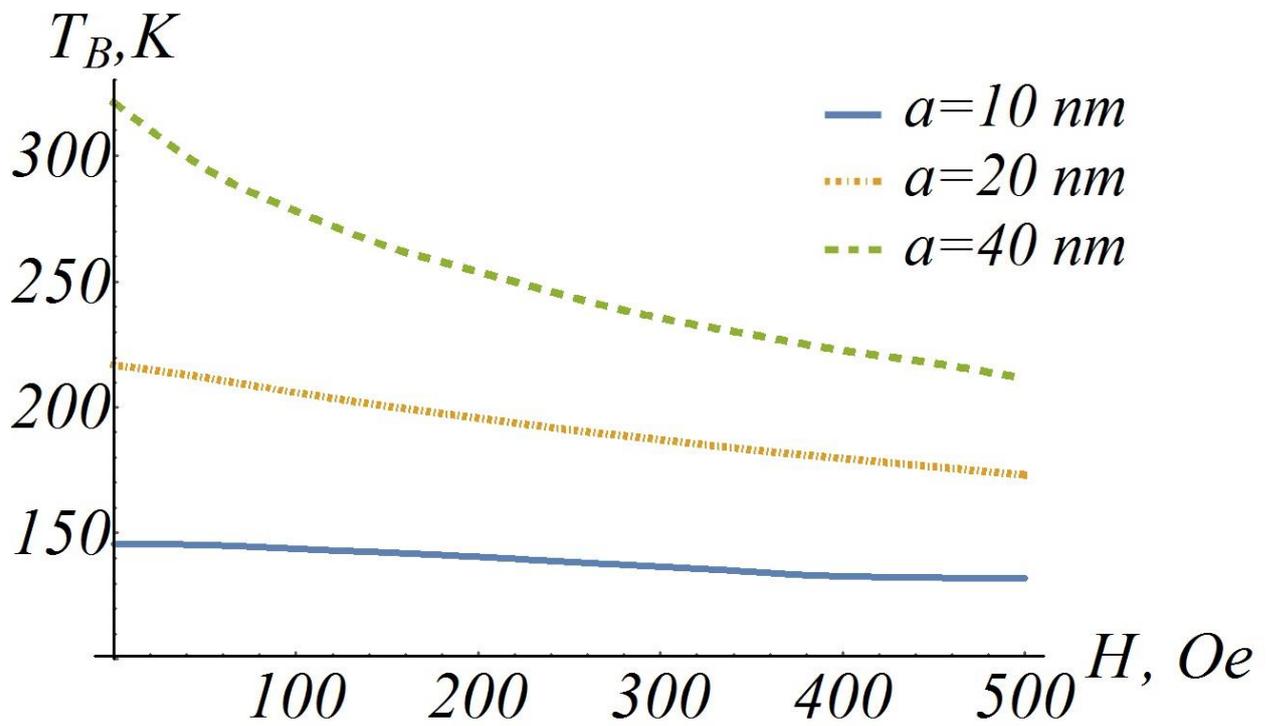

**Fig. 3.** Dependence of the blocking temperature $T_b$ on the external magnetic field H of nanoparticles $Zn_{0.5}Mn_{0.5}Fe_2O_4/Fe_3O_4$ of different sizes.

A decrease in the temperature of blocking the system of nanoparticles with an increase in the magnetic field is associated with a decrease in the barriers of the smallest height. The result obtained is in good agreement with experimental data [5, 21, 24].

### 3.4. Effect of magnetostatic interaction on blocking temperature

The results of modeling the effect of the magnetostatic interaction of $Zn_{0.5}Mn_{0.5}Fe_2O_4/Fe_3O_4$ nanoparticles on their blocking temperature are presented in Fig. 4.

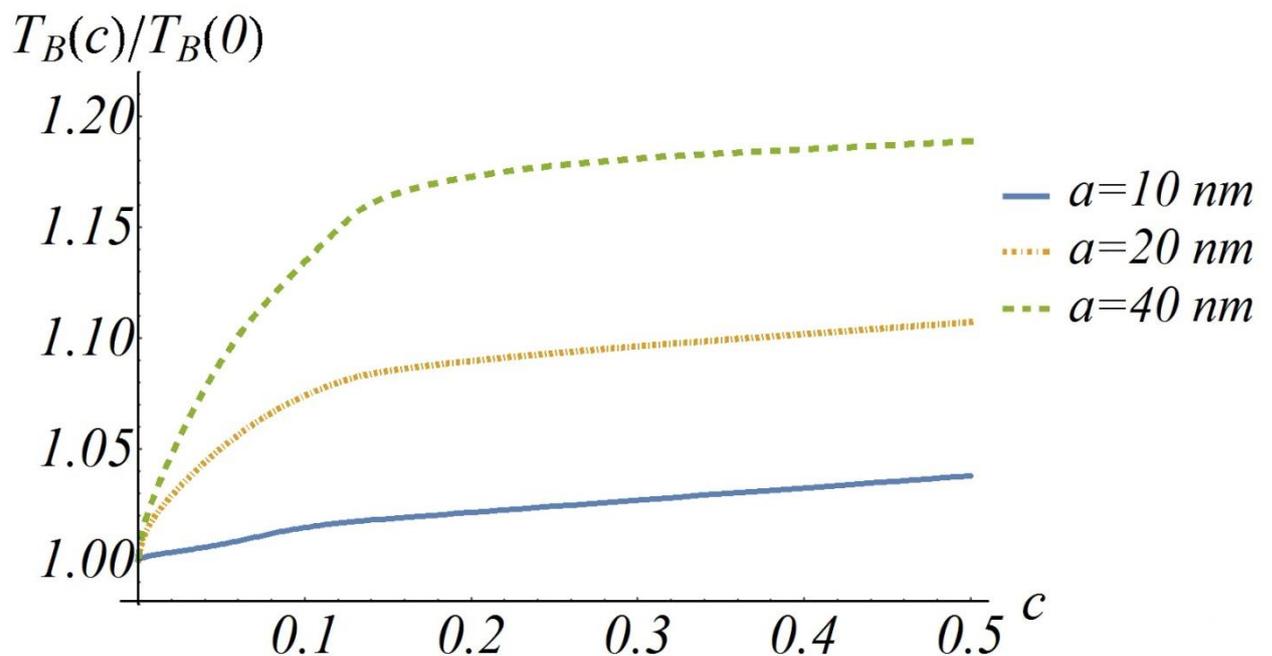

**Fig. 4.** The dependence of the relative blocking temperature $T_b(c)/T_b(0)$ on the volume concentration c of nanoparticles $Zn_{0.5}Mn_{0.5}Fe_2O_4/Fe_3O_4$ of various sizes a. The blocking temperature of non-interacting

particles of size $a = 10$ nm is $T_b(0) = 145,8$ K, for $a = 20$ nm is $T_b(0) = 205.7$ K, and for $a = 40$ nm is $T_b(0) = 324.5$ K.

As the calculation shows, an increase in the interaction intensity (concentration of nanoparticles c) leads to an increase in $T_b$, which is consistent with the results of experimental [5, 11, 13, 38] and theoretical [13, 16, 17, 38, 40, 41] studies. An increase in the blocking temperature with an increase in the concentration of nanoparticles is associated with an increase in the randomization of their magnetic moments [32, 38, 42]. Note that, like in [41], the blocking temperature of large particles ($a = 40$ nm) increases more than in the system of smaller nanoparticles ($a = 10$ nm). For example, at a very high concentration of nanoparticles ($c = 0.5$), $T_b$ of interacting particles changes by 4%, 9%, and 19% compared to $T_b$ of non-interacting particles for particles of 10 nm, 20 nm, and 40 nm, respectively. The noted feature of the behavior of $T_b$ is due to an increase in the magnetic moments of the particles ($m \sim a^3$).

### 4. Calculation of blocking temperature using FC and ZFC curves

The results of modeling the FC and ZFC curves of a system of noninteracting nanoparticles $Zn_{0.5}Mn_{0.5}Fe_2O_4/Fe_3O_4$ of various sizes are presented in Fig. 5. Here, the distribution functions for blocking temperatures are showed, which were determined using the relation:

$$F(T_b)dT_b \sim \left| \frac{1}{[FC(T \to 0) - ZFC(T \to 0)]} \frac{d[FC(T) - ZFC(T)]}{dT} \right|_{T=T_b} dT_b. \quad (8)$$

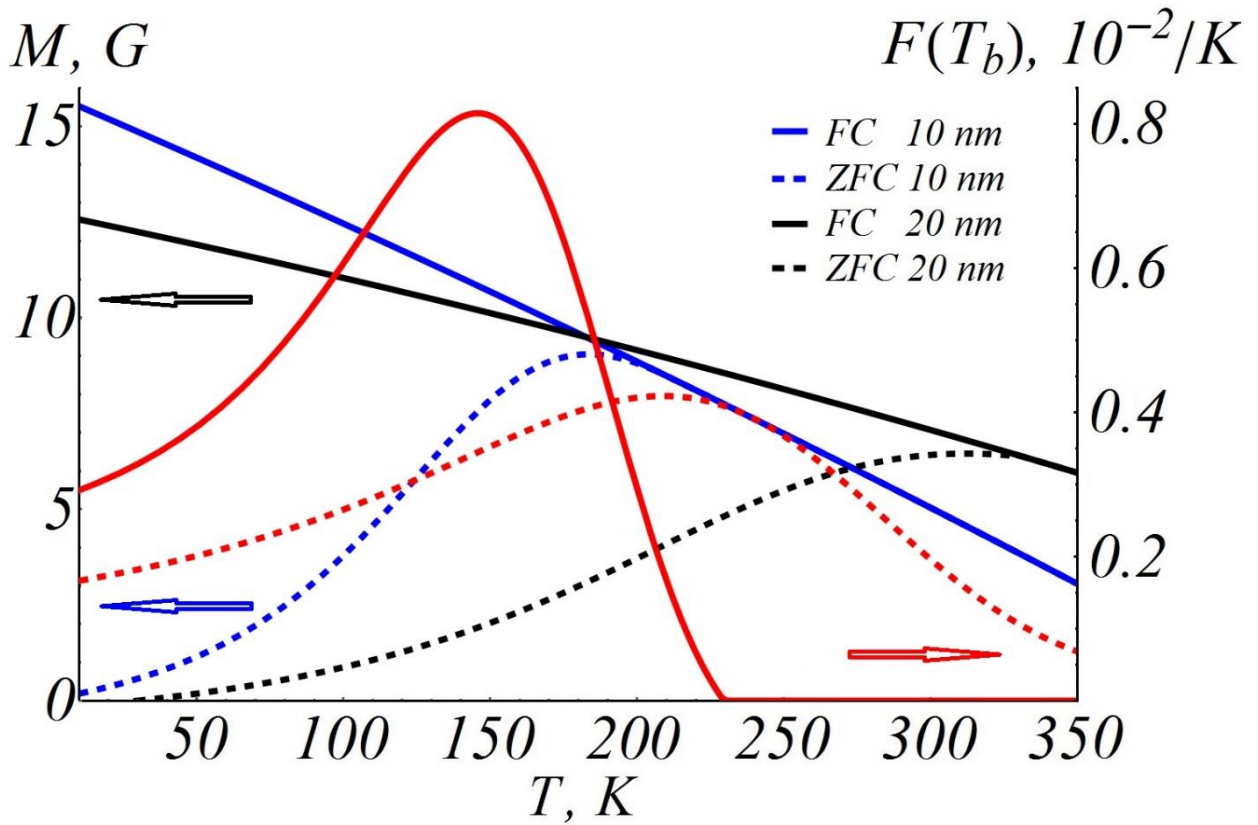

**Fig. 5.** The FC and ZFC curves and the blocking temperature distribution function $F(T_b)$ calculated for $Zn_{0.5}Mn_{0.5}Fe_2O_4/Fe_3O_4$ nanoparticle systems with sizes $\langle a \rangle = 10$ nm and 20 nm. The cooling field H = 100 Oe. Solid red line represents $F(T_b)$ for nanoparticles with size $\langle a \rangle = 10$ nm, whereas dashed red line is for size $\langle a \rangle = 20$ nm.

As expected, with increasing nanoparticle size, the FC-ZFC curves and the corresponding distribution functions $F(T_b)$ shift to higher temperatures. Thus, the blocking temperatures of ensembles of nanoparticles with average sizes $\langle a \rangle = 10$ nm and 20 nm determined by the point of deviation of the ZFC curve from FC, $T_b^{(1)}$, from the maximum of the ZFC curve, $T_b^{(\max)}$ and from the maximum the distribution functions $F(T_b)$, $T_b^{(\text{div})}$, are $T_b^{(1)} = 229$ K and 327 K, $T_b^{(\max)} = 204$ K and 327 K (ZFC curve maximum coincides with split point $T_b^{(1)}$), $T_b^{(\text{div})} = 146$ K and 209 K, respectively. Note that the blocking temperature values calculated for a system of non-interacting nanoparticles of the above sizes using relation (6) $T_b = 144$ K and 206 K are closer to $T_b^{(\text{div})}$.

For comparison with experiment [8], we used various methods to calculate blocking temperatures. The calculation results for $T_b^{(1)}$, $T_b^{(\max)}$, $T_b^{(\text{div})}$ are presented in Table. 2. The table contains the values of $T_b$ obtained using relation (6) for noninteracting particles, as well as the experimental values of $T_b^{(\exp)}$ [8].

**Table 2.** Dimensional dependence of blocking temperature values calculated by the point of deviation of the ZFC curve from FC, $T_b^{(1)}$, by the maximum of the ZFC curve, $T_b^{(\max)}$, by the maximum of the distribution function $F(T_b)$, $T_b^{(\text{div})}$, and using relation (6), $\langle T_b \rangle$. $T_b^{(\exp)}$ are the experimental results [8].

| $\langle a \rangle$, nm | 10 | 11.5 | 12.3 | 13 |
|---|---|---|---|---|
| $T_b^{(1)}$, K | 229 | 230 | 225 | 220 |
| $T_b^{(\max)}$, K | 204 | 207 | 204 | 199 |
| $T_b^{(\text{div})}$, K | 146 | 151 | 143 | 133 |
| $\langle T_b \rangle$, K | 144 | 145 | 140,5 | 135 |
| $T_b^{(\exp)}$, K | 145 ± 5 | 150 ± 5 | 140 ± 5 | 130 ± 5 |

Comparing the methods for calculating the blocking temperature (see Table 2), it is easy to see that the most accurate method is based on differentiating the difference between ZFC and FC magnetizations. The authors of [10] came to a similar conclusion who carried out theoretical and experimental modeling of various methods for determining $T_b$ for single-domain magnetite nanoparticles.

## 5. Conclusion

On the basis of the model of two-phase interacting particles, a method has been developed to estimate the blocking temperature $T_b$ core/shell of nanoparticles based on the calculation of the eigenvalues of the transition matrix. The incorrectness of the estimation of $T_b$ core/shell nanoparticles using the Neel relation was shown.

The size dependence of the blocking temperature $T_b$ of the core/shell nanoparticle system is studied. The "anomalous" dependence of $T_b$ on the sizes of core/shell nanoparticles $Zn_{0.5}Mn_{0.5}Fe_2O_4/Fe_3O_4$ experimentally studied in [8] was simulated. It turned out that a decrease in the blocking temperature with increasing size is associated with a decrease in the height of the main potential barriers separating the magnetic states of $Zn_{0.5}Mn_{0.5}Fe_2O_4/Fe_3O_4$ nanoparticles.

The calculation of the dependence of $T_b$ on the external field $H$ confirmed the decrease in the blocking temperature with increasing $H$ obtained by various authors [5, 21, 24].

Modeling the effect of magnetostatic interaction on $T_b$ showed that an increase in the concentration of nanoparticles leads to an increase in the blocking temperature. Moreover, the $T_b$ of larger nanoparticles changes more significantly, which is due to higher values of magnetic moments. The results obtained are consistent with experimental and theoretical data [5, 11, 13, 16, 17, 38, 40, 41].

From a comparison of different methods for determining the blocking temperature from the ZFC and FC curves, it follows that, as in the case of single-domain nanoparticles [10], differential method is preferred, namely, the maximum difference between ZFC and FC magnetizations.

**Acknowledgement**

This work was financially supported by the state task of the Ministry of Science and Higher Education (Russia) №3.7383.2017/8.9